\documentclass[
    preprint,      
    aps,          
    prx,          
    floatfix,     
    showpacs,     
   nofootinbib   
]{revtex4-2}

\usepackage{graphicx}   
\usepackage{amsmath}    
\usepackage{amssymb}
\usepackage{bm}         

\newcommand{\dd}{\mbox{d}}

\newcommand{\er}{\mathrm{e}}
\newcommand{\bea}{\begin{eqnarray}}
\newcommand{\eea}{\end{eqnarray}}
\newcommand{\be}{\begin{equation}}
\newcommand{\ee}{\end{equation}}

\begin{document}

\title{Restoration of Ensemble Equivalence by Quantum Fluctuations}

\author{Alessandro Campa}
\affiliation{National Center for Radiation Protection and Computational Physics, Istituto Superiore di Sanit\`{a}, Viale Regina Elena 299, 00161 Roma, Italy}
\affiliation{INFN Roma1, I-00185 Roma, Italy}

\author{Andrea Trombettoni}
\affiliation{Department of Physics, University of Trieste, Strada Costiera 11, I-34151 Trieste, Italy}

\begin{abstract}
We study the thermodynamic phase diagram of a one-dimensional quantum spin chain subjected to both mean-field and
nearest-neighbor interactions, and to a transverse magnetic field $h$. The purpose is to determine the effect of the
quantum fluctuations, due to the transverse field, on the phase diagram, in particular with respect to the occurrence of
ensemble inequivalence. We denote our model as a quantum Nagle-Kardar model. To perform the calculation of the canonical partition
function, we show that, due to the presence of the mean-field term, in the thermodynamic limit one can use the Hubbard-Stratonovich
transformation in spite of the non-commutativity of the different operators appearing in the Hamiltonian, and we adopt a
procedure of successive approximations that lead to the determination of
the phase diagram thanks to a scaling property of the phase transition lines. The results show that the ensemble
inequivalence, present in the classical Nagle-Kardar model, is removed above a threshold value $h_c$ for the transverse
field. For $h$ larger than $h_c$ the phase diagram exhibits only second-order phase transition lines,
implying therefore restoration of ensemble equivalence.

\end{abstract}

\maketitle

\section{Introduction}
\label{sec:intro}
Ensemble equivalence is a key result in statistical
mechanics \cite{huang1987statistical,ruelle1999statistical} and it guarantees
that different ensembles give, in the thermodynamic limit, the same result for the macroscopic observables.
A statistical ensemble refers to a collection of systems, identical among them, but in different microscopic states.
For a macroscopic system  we are not interested to (and generally we cannot) have access to the
whole microscopic information, but we aim at determining few macroscopic properties. The latter may depend on the conserved
quantities in the dynamics of the system and on the way it interacts with the environment, typical example being
the microcanonical ensemble (in which the number of particles $N$ and the energy $E$ are strictly fixed) or the canonical
ensemble (in which the number $N$ is fixed, but the system exchanges energy with a bath at a given temperature).

Each ensemble represents specific and different physical situations, related to the control
variables that are employed
for the preparation of the system. A main result of statistical mechanics is that under general assumptions different ensembles give,
for large $N$, the same result for thermodynamic quantities; this is referred to as {\it ensemble equivalence}.
One crucial assumption to have ensemble equivalence is that the interactions among the constituents of the system should
be of finite range or short-range, where the latter definition is here used to define interactions that at large
distance $r$ decay at least as quickly as $1/r^\alpha$, with $\alpha > d$, with $d$ being the dimensionality of the
system \cite{huang1987statistical,ruelle1999statistical}. Ensemble equivalence is no more guaranteed (although
not always violated) in systems where the decay of interaction between the constituents is proportional
to $1/r^\alpha$ with $\alpha \le d$ \cite{campa2009statistical,campa2014physics,defenu2023longrange}. Systems, either classical or quantum,
with this characteristic, can be denoted as long-range systems, that thus can present the phenomenon
of {\it ensemble inequivalence}. A much studied example, especially in the case of magnetic systems, is the one of
mean-field systems, where the parameter $\alpha$ above is equal to $0$, and it corresponds to an all-to-all
interaction.
It is useful to remark that the inequivalence is connected to the fact that
the equilibrium state of the system depends on its preparation, or, in other words, on which thermodynamic
quantities are taken as control variables.

Inequivalence of ensembles has been studied in detail in classical long-range systems \cite{barre2001inequivalence,
choi2003stability,mukamel2005breaking,bouchet2005classification,kastner2006topological,campa2006dynamics,dauxois2010models,
chavanis2011thermodynamics,teles2012nonequilibrium,pikovsky2014ensemble,chakhmakhchyan2017ensemble,hovhannisyan2017complete,
rochafilho2018ensemble,prasad2019ensemble,campa2019ising,baldovin2021physical,hou2021zeroth,gray2022electric,
yang2024ensemble,campa2025ensemble,evnin2025ensemble},
including long-range systems in presence of randomness \cite{bertalan2011ensemble,bertalan2011ensemble-2,murata2012ensemble}
and the semi-classical description of cold atoms interacting via photon-mediated
long-range forces \cite{keller2017phases}. Generally, having only a single
long-range ferromagnetic interaction, even of
the mean-field type, is not enough to give rise to ensemble inequivalence, that instead
is typically found if in
addition there are also multi-spin interactions \cite{bertalan2011ensemble,prasad2019ensemble} or,
like in the spin system now known as the Nagle-Kardar model \cite{nagle1970ising,kardar1983crossover}, there is
a short-range term competing with the long-range interaction \cite{mukamel2005breaking,campa2025ensemble}. We remind that
when a nearest-neighbor interaction is present together with a ferromagnetic mean-field interaction
\cite{nagle1970ising,kardar1983crossover,baker1963ising,bonner1971ising},
one has at $T=0$ a change in the ground-state from a ferromagnetic to an anti-ferromagnetic one at a
negative value of the ratio $K/J$, where $J>0$ is the strength of
the all-to-all coupling and $K$ is the nearest-neighbor coupling:
the competition between the two types of interactions produces a phase transition a $T=0$.

Recent works addressed the ensemble inequivalence in quantum long-range systems
\cite{russomanno2021quantum,defenu2024ensemble,arrufatvicente2025ensemble}. In \cite{russomanno2021quantum} a
quantum Ising chain with long-range interactions of the form $S_i^z S_j^z$
was studied in presence of an transverse field $\propto \sum_i S_i^x$ (with the $S_{x,z}^i$ referring to the
$x,z$ Pauli matrices of the spin in the $i$-th position of the chain),
arguing that for $\alpha < 1 = d$ the microcanonical entropy
is nonconvex, as a consequence of the fact the spectrum
is organized in energetically separated multiplets, a property typical of
quantum systems in their long-range regime \cite{defenu2021metastability}. In
\cite{defenu2024ensemble,arrufatvicente2025ensemble} the ensemble inequivalence was shown to occur
at finite temperature in a model having all-to-all $2$- and $4$- spin interactions [respectively of the
form $(\sum_i S_i^z)^2$ and $(\sum_i S_i^z)^4$] also in presence of a transverse field].

In this paper we address the issue of whether and how ensemble equivalence
can be restored by quantum fluctuations. Since ensemble
inequivalence is taking place in both classical and quantum systems, a general question is what is the effect on it if
one turns on quantum fluctuations on a classical long-range model featuring inequivalence.
Denoting by $h$ the parameter characterizing the strength of the term giving rise to the quantum fluctuations (such as
a transverse field), in general one may expect
that the degeneracies typical of systems in the long-range regime are spoiled or removed.
Then the possible presence of a critical value $h_c$ to {\it restore}
ensemble equivalence may be foreseen.
This argument relies on the assumption that, as far as the order properties of the equilibrium state
are concerned, quantum fluctuations might play a role similar to that of thermal fluctuations. This assumption
could be considered together with the following known fact. The presence of
first-order phase transitions in a long-range system implies inequivalence
of the canonical ensemble with the microcanonical ensemble \cite{campa2014physics}. In fact,
the free energy density is the Legendre-Fenchel transform of the microcanonical entropy and as such it is a concave function.
A discontinuity in it, i.e., a first-order phase transition, implies the presence of a convex part in the
microcanonical entropy. In short-range systems this convex part is substituted by a linear section, related physically
to phase separation, thus restoring equivalence. In long-range systems there is no phase separation, and then one has
ensemble inequivalence. One also observes that in general a system presents first-order transitions at lower
temperatures with respect to second-order transitions (this being obviously a general trend, not a rigorous fact).
Then, if quantum fluctuations play a role similar to thermal fluctuations for the determination of the equilibrium
state, one can reach the conclusion, stated above, of the presence of a critical value $h_c$
above which first-order transitions do not occur even at very low temperature.

We remark also the following important methodological issue. The property that the presence
of first-order transitions in the canonical ensemble implies inequivalence with the microcanonical ensemble, allows to
infer the occurrence of inequivalence without actually performing calculations in the latter ensemble. Of course,
these calculations would be necessary to determine quantitatively the thermodynamic phase diagram, but in this
work our aim is only to see when inequivalence occurs, and if and when ensemble equivalence is restored.

To perform our study, we have looked for a system such that: {\it i}) the interactions are not
purely mean-field, but there are also local (actually nearest-neighbor) interactions; {\it ii}) in the classical
limit ($h=0$) it exhibits ensemble
inequivalence; {\it iii}) it is treatable, having a phase diagram which can be determined also in presence of competing
interactions. The paper is organized as follows. In Section \ref{sec:model} we introduce our model, we remind its
equilibrium properties in the various limits (i.e., in the classical limit, in the limit without nearest-neighbor
interaction, in the limit without mean-field interaction), and we give the first description of our method of
computation. The method is then explained in details in Section \ref{sec:method}. In Section \ref{sec:results} we
show our results for the thermodynamic phase diagram and the implications for ensemble equivalence or inequivalence.
Section \ref{sec:concl} presents the discussion and the conclusion. Some technical issues are relegated to the Appendices.

\section{The model}
\label{sec:model}
It is very difficult to have a model
satisfying all the above three requests {\it i}), {\it ii}) and {\it iii}). We decided to study a model
satisfying the first two, while its treatability relies on a new technique that we introduce and that it is based on successive
approximations; the technique is explained in details in the next Section.
The model we have chosen is a $1D$ spin chain that has the following three parameters:
$J>0$, the mean-field ferromagnetic coupling (we remind that if $J<0$
there cannot be an antiferromagnetic order at finite temperature in $1D$); 
$K$, the nearest-neighbor coupling between the spins; and
$h$, the transverse magnetic field. The Hamiltonian then reads:
\be
H = -\frac{J}{2N} \sum_{i,j} S_i^z S_j^z - \frac{K}{2} \sum_i S_i^z S_{i+1}^z -h \sum_i S_i^x \, ,
\label{origham}
\ee
where $N$ is the number of spins. We assume periodic boundary conditions, $S_{N+1}^{x,z}=S_1^{x,z}$; as already mentioned above,
$S_{x,z}^i$ are the $x,z$ Pauli matrices. The normalization with $N$ of the coupling coefficient of the mean-field term
is the usual Kac prescription adopted in long-range systems; it makes the systems extensive, obviously without
removing the non-additivity, the feature responsible for ensemble inequivalence \cite{campa2009statistical,campa2014physics}.
It is possible to consider only nonnegative values of $h$, since the Hamiltonian obtained changing $h$ in $-h$ is
unitarily equivalent. 


All the three limits of the model are very well studied.

{\it a}) For $h=0$ (no quantum fluctuations) it reduces
to the classical Nagle-Kardar model \cite{nagle1970ising,kardar1983crossover}, which is solvable and it 
exhibits ensemble inequivalence in the region where $K/J$ is negative
\cite{mukamel2005breaking}. The phase diagram in the $\left(K/J,T/J\right)$ plane 
features a first-order phase transition line at low temperature, that starts at $T=0$ at the point
$(K/J) = -0.5$, and ends at the tricritical point $(K/J)_{tr} = - \ln \sqrt{3}/\sqrt{3} \approx -0.3171$, $(T/J)_{tr} = 1/\sqrt{3} \approx 0.5774$.
The phase transition is between a paramagnetic and a ferromagnetic state. At the tricritical point the phase transition line, still separating
a paramagnetic and a ferromagnetic state, becomes second-order, and it is implicitly defined by $[(T/J)_c]^{-1} = \exp[ -(K/T)_c]$. This expression shows that
the critical temperature $T_c$ monotonously increases with $K/J$, and it is $T_c=J$ for $K=0$.

{\it b})
For $J=0$ the model reduces to the quantum Ising chain
in a transverse field \cite{sachdev2011quantum,chakrabarti2008quantum}, which is solvable by the Jordan-Wigner transformation, that shows that the spin system
can be mapped to a collection of independent fermions. It features a second-order quantum phase
transition at $T=0$, and no transition for $T>0$.

{\it c})
For $K=0$ the model reduces to the Lipkin-Meshkov-Glick (LMG) model
\cite{lipkin1965validity,lipkin1965validity-2}, a well studied system with all-to-all interactions in presence of a transverse field, having 
a spectrum that can be determined analytically
\cite{pan1991analytical,ribeiro2008exact}. It has a second-order phase transition both at $T=0$ and at finite temperature. 

Due to its different limits,
we can refer to this model as a {\it quantum} Nagle-Kardar model,
or as a quantum Ising chain with added mean-field interactions,
or to as an {\it Ising-}LMG. Given that our motivation is to study ensemble
inequivalence and the Nagle-Kardar model is a paradigmatic classic model
featuring it, we will use the first of these definitions. 

Dynamical phase transitions of this model 
have been studied in \cite{lerose2018chaotic,lerose2019impact} where the model was introduced, quenching $h$ from zero to a finite value with $J$ and $K$ positive. Other
quantum models with 
both short- and long-range interactions have been also discussed 
\cite{igloi2019quantum,belyansky2020minimal,lerose2020origin,granet2023exact,zunkovic2024meanfield}, but to the best of our knowledge ensemble (in)equivalence has been not addressed. 
The point is that to have ensemble inequivalence in models such as the Nagle-Kardar model one needs competing long- and short-range interactions and to study the quantum
Nagle-Kardar model with $K/J$ negative (and $J>0$) is very challenging due to the competition between ferromagnetic long-range coupling and anti-ferromagnetic short-range
ones in presence of quantum fluctuations, in particular near the point
$(K/J)_c$ at which the classical Nagle-Kardar model exhibits a (ground-state) phase transition at $T=0$. The reason for such difficulty is that standard methods 
such as spin-wave theory \cite{auerbach1994interacting,manousakis1991spin}
require to do linearization around a well-defined (ferro- or anti-ferromagnetic) ground-state and become 
more involved where the two compete. Also, the Jordan-Wigner transformation, that is useful for other models with short- and long-range interactions
\cite{igloi2019quantum,granet2023exact}, here is not of help for the quantum Nagle-Kardar model,
in that it does not give rise to a solvable fermionic model.

In classical systems a general method to deal with spin models using successive approximations is that know as Bethe-Peierls approximation.
It uses a cluster mean-field computation. In short (see however Appendix \ref{app:bethe} for a specific and more detailed application to our model in the classical
limit $h=0$), in presence of a short-range model, in this method one solves exactly a model of only $p$ spins, the Hamiltonian of which contains also parameters related
to the effect of the rest of the system; the parameters are determined self-consistently
\cite{pathria_beale2021statistical,oguchi1953theory,yamamoto2009correlated,strecka2015brief}. The model with only one spin, $p=1$, corresponds to the mean-field approximation,
and it is expected that for increasing $p$ one approaches the exact solution for the original model. Generally, for a $1D$ chanin, the step beyond the mean-field approximation consists
of choosing $p=3$, so that the nearest-neighbor of the central site are included \cite{pathria_beale2021statistical}. The method can be extended also
to models with a mean-field interaction. In Appendix \ref{app:bethe} we show that, remarkably, in the classical limit of our model, i.e., in the classical
Nagle-Kardar model, the Bethe-Peierls approximation with $p=3$ produces already the {\it exact} result for the phase diagram.

Things are different for the quantum case, when $h$ is non-vanishing. One can see that the Bethe-Peierls approximation, even at larger values of $p$
does not produce the exact result already at $T=0$ for the quantum Ising chain in a transverse field. Our strategy then will be different. It is
a method that could still be seen in the framework of a cluster approximation using increasing values of $p$, but that does not include effective parameters to be
optimized. The strenght of the method resides in the fact that, at least for the most relevant region of the thermodynamic phase diagram, the region where the
first-order phase transitions and ensemble inequivalence are located, a simple rescaling of the results (a procedure explained in details below) found
at relatively small values of $p$, allows to obtain the results pertaining to the original model, formally approached for $p\to \infty$.
This represents a major result of this paper, in that one can find the phase transition lines of
the quantum Nagle-Kardar model, and in particular first-order lines as a function of $h$, allowing to describe how restoration
of ensemble inequivalence by quantum fluctuations takes place.

\section{Method of calculation}
\label{sec:method}

Although not directly related to our method for the quantum Nagle-Kardar model, we find it instructive and useful to present, in Appendix \ref{app:bethe},
the Bethe-Peierls approximation that can be employed in the classical limit of our Hamiltonian (\ref{origham}). The main interest resides in the fact that for
this model the approximation with $p=3$ already gives the exact result. The latter can be obtained with the help of the Hubbard-Stratonovich (HS) transformation,
which is based on the Gaussian identity
\be
\label{hstrans}
\er^{ba^2} = \sqrt{\frac{b}{\pi}} \int \dd x \, \er^{-bx^2 +2abx} \, ,
\ee
where $a$ and $b$ are real numbers with $b>0$.

\begin{figure}[htbp]
\begin{center}
 \includegraphics[clip, trim=0cm 0cm 0cm 0cm, width=0.70\textwidth]{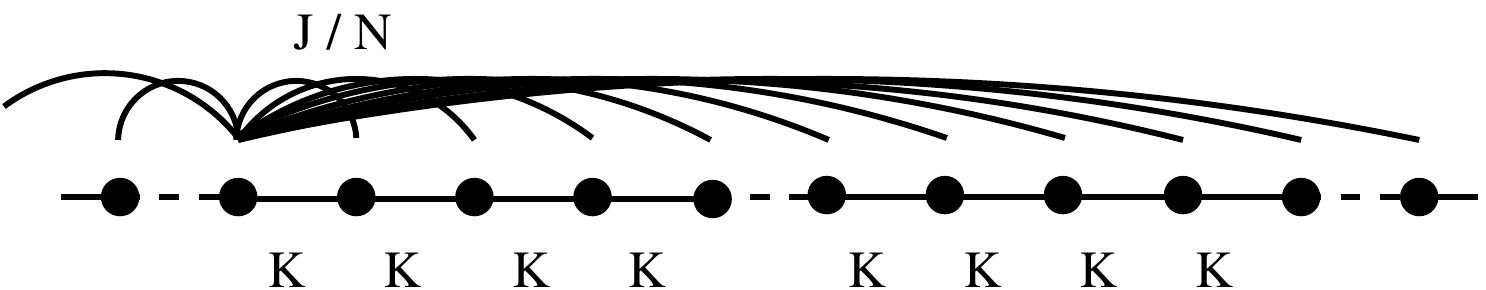}
\caption {The clustering adopted for our calculations. The system is still fully connected through the mean-filed interaction proportional to $J/N$
(in the picture this interaction is actually shown only for the second spin from the left). On the other hand, the nearest-neighbor interaction proportional to $K$
is removed for one every $p$ couples; in the picture the couples for which this interaction is removed are joined by a dashed line, rather than a full line.}
\label{fig1}
\end{center}
\end{figure}

Although inspired by the Bethe-Peierls approximation for the classical limit, we will perform the clustering in the quantum case in a different way, that is explained
in details in this Section. We anticipate that we can talk of
a clustering of the system since our approximation consists in computing the partition function for
a system in which one every $p$ couplings $K$ between nearest-neighbors is removed (see Fig. \ref{fig1} for a visual representation).
It is important to note that, even after this removal,
the system is still fully connected, due to the presence of the mean-field interaction proportional to $J/N$ (notice that the $J$ term is typically not present
on studies of Ising models with cluster
mean-field techniques \cite{yamamoto2009correlated,strecka2015brief}). However, in the effective
Hamiltonian defined below, see Eqs. (\ref{Hpeff}) and (\ref{Hpeffs}), the
system decouples in blocks of $p$ spins. We have performed the calculations for various values of $p$, that we specify presenting the results. Obviously the exact Hamiltonian
is recovered when $p \to \infty$. The reason to introduce the cut
of the nearest-neighbor interaction every $p$ spins is that the computation becomes feasible and that, more importantly, from the results obtained
for the computationally accessible values of $p$ one can infer the behavior for $p \to \infty$ in the most interesting part of the phase diagram. This will be seen very clearly
below. But first we want to explain in some details the rationale behind our method of computation.

The two main ingredients of the method are: {\it A}) the HS transformation; {\it B}) as just anticipated, the approximation of the Hamiltonian with one in which the
nearest-neighbor interaction is removed every $p$ spins. It is important to explain the reason for the choice of this method and the physical and mathematical issues
related to it. Let start from point {\it A}). The HS transformation uses the Gaussian identity (\ref{hstrans}) introducing an auxiliary variable, and physically it can be
interpreted as the replacement of the mean-field interaction, that couples each spin with all the others, with an interaction of each spin with an additional
longitudinal magnetic field. In the quantum case, i.e., when $h \ne 0$, this replacement is actually not exact, or, in other words, for a finite system the partition function
computed with the use of the HS transformation is not equal to the partition function of the original system. This can be seen as follows.
The partition function of the system is 
\be
\label{partfull}
Z = {\rm Tr}\, \er^{-\beta H} \, ,
\ee
where $H$ is the Hamiltonian given in Eq. (\ref{origham}), and where the trace ${\rm Tr}$ is on the spin variables $S$'s. 
With the HS transformation, one actually considers the partition function
\be
\label{parths}
\widetilde{Z}=\sqrt{\frac{\beta N}{2 \pi}} 
\int \dd x \, {\rm Tr} \exp \left\{ -\frac{N\beta}{2}x^2 \right\}
\exp \left\{ -\beta \widetilde{H}(x) \right\} \,
\ee
where the effective Hamiltonian $\widetilde{H}(x)$ is given by
\be
\label{tildeham}
\widetilde{H}(x) = -\frac{K}{2} \sum_{i} S_{i}^z S_{i+1}^z -h \sum_{i} S_{i}^x
-\sqrt{J}x \sum_{i} S_{i}^z \, .
\ee
The problem is that $Z$ and $\widetilde{Z}$ are ``in general'' not equal, due to the non-commutativity of the operators $S_{i}^x$'s and $S_{i}^z$'s; indeed, the exponential of
$\widetilde{H}(x)$ cannot be written as the exponential of the part of $\widetilde{H}(x)$ with the $S_{i}^z$'s (let us call it $\widetilde{H}_z$) times the exponential of the part of
$\widetilde{H}$ with the $S_{i}^x$'s (let us call it $\widetilde{H}_x$), due to the fact the $S_{i}^x$'s do not commute with the $S_{i}^z$'s. In other words,
it is $\widetilde{H}=\widetilde{H}_z+\widetilde{H}_x$, but $\er^{-\beta \widetilde{H}} \neq \er^{-\beta \widetilde{H}_z} \, \er^{-\beta \widetilde{H}_x}$ since
$\left[ \widetilde{H}_z,\widetilde{H}_x \right] \neq 0$. This problem would not arise in the classical case, i.e., for $h=0$, since then $\widetilde{H}_x=0$
and we would have  equality of $Z$ and $\widetilde{Z}$. Nevertheless, in the quantum case with $K=0$, i.e., in the LMG model reminded above, it is argued that
despite the non-commutativity of $\widetilde{H}_z$ and $\widetilde{H}_x$, it is anyway $\widetilde{Z}=Z$
{\it in the thermodynamic limit}, as described in Ref. \cite{romanroche2023exact}.

Actually, one can use an argument to show that also in our case, in which $K\ne0$, in the thermodynamic limit it is possible to
neglect the commutator and thus employ the HS transformation. The argument thus includes also the case with $K=0$.
To this purpose, let us suppose to have two operators, denoted by $A$ and $B$, that do not commute, and let us write the expression
\be
\label{arg1}
\int \dd x \, \er^{-N \frac{x^2}{2}} \er^{A+xB}
\ee
We would like that for $N\to \infty$ this would be equal, apart from a multiplicative factor, to
\be\label{arg2}
\er^{A+\frac{1}{2N}B^2}
\ee
This equality would suffice to show that in the thermodynamic limit $Z$ in Eq. (\ref{partfull}) and $\widetilde{Z}$ in Eq. (\ref{parths}) are equal, by taking
$A \to \beta K \sum_{i} S_{i}^z S_{i+1}^z +\beta h \sum_{i} S_{i}^x$ and $B \to \beta \sqrt{J}x \sum_{i} S_{i}^z$ (and, for the sake of precision, $N\to \sqrt{\beta}N$).
The expression in (\ref{arg1}) can be rewritten, using the Trotter formula, as
\be
\label{arg3}
\int \dd x \, \er^{-N \frac{x^2}{2}} \lim_{n\to \infty}\left( \er^{\frac{A}{n}}\er^{\frac{xB}{n}}\right)^n
\ee
For a given $n$ this expression could be written also as
\be
\label{arg4}
\int \dd x \dd y_1 \dots \dd y_n \, \delta(x-y_1)\dots \delta(x-y_n)
\er^{-\frac{N}{n} \frac{y_1^2+\dots +y_n^2}{2}}
\left( \er^{\frac{A}{n}}\er^{\frac{y_1 B}{n}}\right)
\left( \er^{\frac{A}{n}}\er^{\frac{y_2 B}{n}}\right)\dots\left( \er^{\frac{A}{n}}\er^{\frac{y_n B}{n}}\right)
\ee
Putting the numerical factor in the first exponential involving $y_i^2$ together with the corresponding
$i$-th factor in the exponential with the operators, we obtain
\be
\label{arg5}
\int \dd x \dd y_1 \dots \dd y_n \, \delta(x-y_1)\dots \delta(x-y_n)
\left( \er^{\frac{A}{n}}\er^{-\frac{N}{2n}y_1^2}\er^{\frac{y_1 B}{n}}\right)
\dots
\left( \er^{\frac{A}{n}}\er^{-\frac{N}{2n}y_n^2}\er^{\frac{y_n B}{n}}\right)
\ee
At this point we can invoke the fact that, for each $y_i$, assuming that $N/n$ is very large (in other words,
assuming that while both $N$ and $n$ tend to infinite, $N$ is always much larger than $n$), the factor
involving $y_i$ is dominated by the value of $y_i$ equal to $B/N$; equivalently, the exponential that we obtain
in this factor by putting $y_i$ equal to $B/N$ is equal to the exponential that we would get by integrating freely
on $y_i$. Contrary to the original expression (\ref{arg1}),
we can invoke the dominance of this value of $y_i$ since here $(A/n)$ and $(y_iB/n)$ appear in different
exponents, while in Eq. (\ref{arg1}) $A$ and $xB$ are in the same exponent, and we could not invoke right away the
dominance of the value $x=B/N$
(the loose notation $x=B/N$, in which a scalar is equalized to an operator, can be formalized more precisely; see Appendix \ref{app:gauss}).
It is true that each $y_i$ is forced, by the Dirac delta, to be equal to $x$, so that we cannot integrate freely on each
$y_i$, but the integration on $x$ allows formally each $y_i$ to explore the all range;
furthermore, since we find that the dominant contribution occurs for a value of $y_i$ which is equal for each $i$,
we do not violate the constraint imposed by the Dirac deltas. In conclusion, Eq. (\ref{arg5}) can be taken to approach,
in the assumptions made above for $N$ and $n$,
\be
\label{arg6}
\left( \er^{\frac{A}{n}}\er^{\frac{1}{2Nn}B^2}\right)^n
\ee
But using again the Trotter formula, this is equal, when $n$ is very large, to
\be
\label{arg7}
\er^{A+\frac{1}{2N}B^2}
\ee
which is what we wanted to obtain. Thus, we can use Eq. (\ref{parths}) to compute our partition function in the thermodynamic limit (we remind that in the classical case,
$h=0$, the HS transformation is ``harmless'' not only in the thermodynamic limit, but also for finite $N$).
Thus, from now on, while we will distinguish the
original Hamiltonian $H$ and the effective $x$-dependent Hamiltonian $\widetilde{H}(x)$, we remove the tilda from the partition function computed in Eq. (\ref{parths}),
identifying it with the partition function $Z$ defined in Eq. (\ref{partfull}).

We see that without the nearest-neighbor interaction (the LMG model, $K=0$) the effective Hamiltonian $\widetilde{H}(x)$ represents a system of independent spins (or better,
a system where the interaction among the spins is mediated through the additional longitudinal field). In this case the partition function $\widetilde{Z}$
in the thermodynamic limit is easily computed with a saddle point evaluation of the integral over the longitudinal field $x$. The thing is different when $K\ne 0$, since in
this case not even in the effective Hamiltonian $\widetilde{H}(x)$ the spins are decoupled, because of the nearest-neighbor interaction. This does not constitute a problem in
the classical case ($h=0$): although the spins are still directly coupled through this interaction, a transfer matrix computation can be employed. On the other hand,
in the quantum case ($h\ne 0$) it is not possible to efficiently compute ${\rm Tr}\, \er^{-\beta \widetilde{H}(x)}$ with a transfer matrix method, since the transfer matrix
would be $2^N \times 2^N$, i.e., of the same computational complexity of a direct diagonalization of the initial Hamiltonian, and we are interested in very large $N$.

We then come here to the point {\it B}) mentioned above, i.e., the approximation of
the Hamiltonian with one in which the
nearest-neighbor interaction is removed every $p$ spins.
One obtains in $\widetilde{H}(x)$ independent blocks
of $p$ spins, so that ${\rm Tr}\, \er^{-\beta \widetilde{H}(x)}$ for a system of $N$ spins, assuming $N$ to be a multiple of $p$, is equal to the $(N/p)$-th
power of the trace restricted to a single block. For $p$ not very large the partition function of the single block can be computed numerically. Then, as for the classical case,
a saddle point evaluation of the integral over the longitudinal field allows to obtain the partition function in the thermodynamic limit. We will see the usefulness of this
approach when presenting the results, by showing that from the results for the computationally accessible values of $p$ one can infer the behavior for $p \to \infty$ in the
most interesting part of the phase diagram. 

In conclusion, the effective $x$-dependent Hamiltonian $\widetilde{H}(x)$, obtained with the HS transformation, is approximated by the following
$x$-dependent and $p$-dependent ($p$-blocking) effective Hamiltonian:
\be
\label{Hpeff}
\widetilde{H}^{(p)}(x) = \sum_s \widetilde{H}_s^{(p)}(x)
\ee
where $\widetilde{H}_s^{(p)}(x)$ denotes the effective Hamiltonian corresponding to the $s$-th block.
Denoting with $S_{i,s}^x$ and $S_{i,s}^z$, $i=1,\dots,p$, the operators belonging to the $s$-th block, we have
\bea
\widetilde{H}_s^{(p)}(x) &=& - \sqrt{J} x \sum_{i=1}^{p}  S_{i,s}^z -\frac{K}{2} \sum_{i=1}^{p-1} S_{i,s}^z S_{i+1,s}^z \nonumber \\
&& -h \sum_{i=1}^{p} S_{i,s}^x \, .
\label{Hpeffs}
\eea
Therefore the partition function $Z$ in Eq. (\ref{parths}) (as pointed out above, identified by now, in the thermodynamic limit, with the exact partition function)
is approximated by the $p$-dependent partition function $Z^{(p)}$ given by
\be
\label{parthblock}
Z^{(p)} = \sqrt{\frac{\beta N}{2 \pi}}\!\! \int \dd x  {\rm Tr} \exp \left\{ -\frac{N\beta}{2}x^2 \right\}
\exp \left\{ -\beta \widetilde{H}_s^{(p)}(x) \right\} \, .
\ee
Notice that we are doing the $p$-blocking after the HS transformation, but one could equivalently approximate the original Hamiltonian $H$ with the $p$-blocking, and only
at that point perform the HS transformation. Namely, we could define the $p$-blocking Hamiltonian
\be
\label{pblockham}
H^{(p)} = \sum_s H_s \, ,
\ee
with
\bea
H_s &=& -\frac{J}{2N} \sum_{s'\ne s} \sum_{i=1}^{p}\sum_{j=1}^p S_{i,s}^z S_{j,s'}^z -\frac{J}{2N}
\sum_{i,j=1}^{p} S_{i,s}^z S_{j,s}^z \nonumber \\
&&-\frac{K}{2} \sum_{i=1}^{p-1} S_{i,s}^z S_{i+1,s}^z -h \sum_{i=1}^{p} S_{i,s}^x \, .
\label{Hcut}
\eea
Obviously the sum over $s$ of the first two terms of $H_s$ is equal to the first term in Eq. (\ref{origham}). Performing now the HS transformation we would obtain
the partition function (\ref{parthblock}), with $\widetilde{H}^{(p)}$ defined in Eqs. (\ref{Hpeff}) and (\ref{Hpeffs}).

If we now define
\be
\label{partialpart}
\exp \left[ -\beta \widetilde{f}^{(p)}(\beta x, \beta K, \beta h) \right] =
{\rm Tr} \exp \left\{ -\beta \widetilde{H}_s^{(p)}(x) \right\} \, ,
\ee
where the trace is over the spins of the block (obviously it does not depend on which block it is performed), and where we have explicitly indicated the dependence
on the inverse temperature $\beta$, the Hamiltonian parameters $K$ and $h$, and the auxiliary field $x$, then we have
\be
Z^{(p)} = \!\!\sqrt{\frac{\beta N}{2 \pi}}\!\! \int \!\! \dd x 
\exp \left\{\!\! -N \!\! \left[\frac{\beta}{2} x^2+\frac{\beta}{p} \widetilde{f}^{(p)}(\beta x,\beta K,\beta h)\right]\right\},
\label{zetacomp}
\ee
Finally, computing the above integral with the saddle point method, we obtain that, for a given $p$ the free energy per spin of our system, in the thermodynamic limit,
is given by: 
\be
\label{free_en_p}
\beta f_p(\beta K, \beta h) = \min_x \left[ \frac{\beta}{2} x^2 + \frac{1}{p} \beta \widetilde{f}^{(p)}(\beta K,\beta h) \right] \, .
\ee

The disordered ground state of the system, for $K<0$ and sufficiently large in modulus, will have vanishing average  magnetization per spin
$\langle \left( \sum S_i^z \right)/N \rangle$; therefore in our calculations we consider only even $p$, so that the disordered ground state of a block will have
vanishing magnetization.

As mentioned above, if $K=0$ it is not necessary to introduce the cut every $p$ spins, and the free energy per spins can be computed easily. This is done
in Appendix \ref{app:lmg}.

\section{Results for the phase diagram}
\label{sec:results}

The value of $x$ that realizes the minimum in Eq. (\ref{free_en_p}) is equal to the average magnetization per spin  $\langle \left( \sum S_i^z \right)/N \rangle$.
As it occurs in models with the symmetry $S_i^z \to -S_i^z$, if there is a positive value of $x$ realizing the minimum, also the value $-x$ realizes it, so from
the mathematical point of view the system magnetizes if we consider that it is put in an infinitesimal external magnetic field in the $z$ direction, and that
this field is sent to zero after the thermodynamical limit is taken. Therefore, to study the phase transitions it is sufficient to study the minimization problem
(\ref{free_en_p}) for non-negative $x$.
It is clear that, unlike the original model ($p=\infty$), the value of $\langle S_i^z \rangle$ is not uniform in $i$. To be more precise, using the above notation
$S_{i,s}^z$ for the $i$-th spin in the $s$-th block ($i=1,\dots,p$), the value of $\langle S_{i,s}^z \rangle$ will depend on $i$ but, for symmetry, it will not
depend on $s$. The dependence on $i$ will vanish for $p \to \infty$. However, we are not concerned with the inhomogeneity inside a block, since the phase
transition, be it first-order or second-order, will be related to the passage of the value of $x$ realizing the minimum form a positive value to zero.
On physical grounds, as can be seen by redefining the unit of energy, the results, when $J\ne 0$, depend only on the combination $K/J$, $h/J$ and $T/J$; therefore
for convenience in the remaining part of the main text we take $J=1$.

\begin{figure}[htbp]
\begin{center}
 \includegraphics[clip, trim=0cm 9cm 0cm 9cm, width=0.70\textwidth]{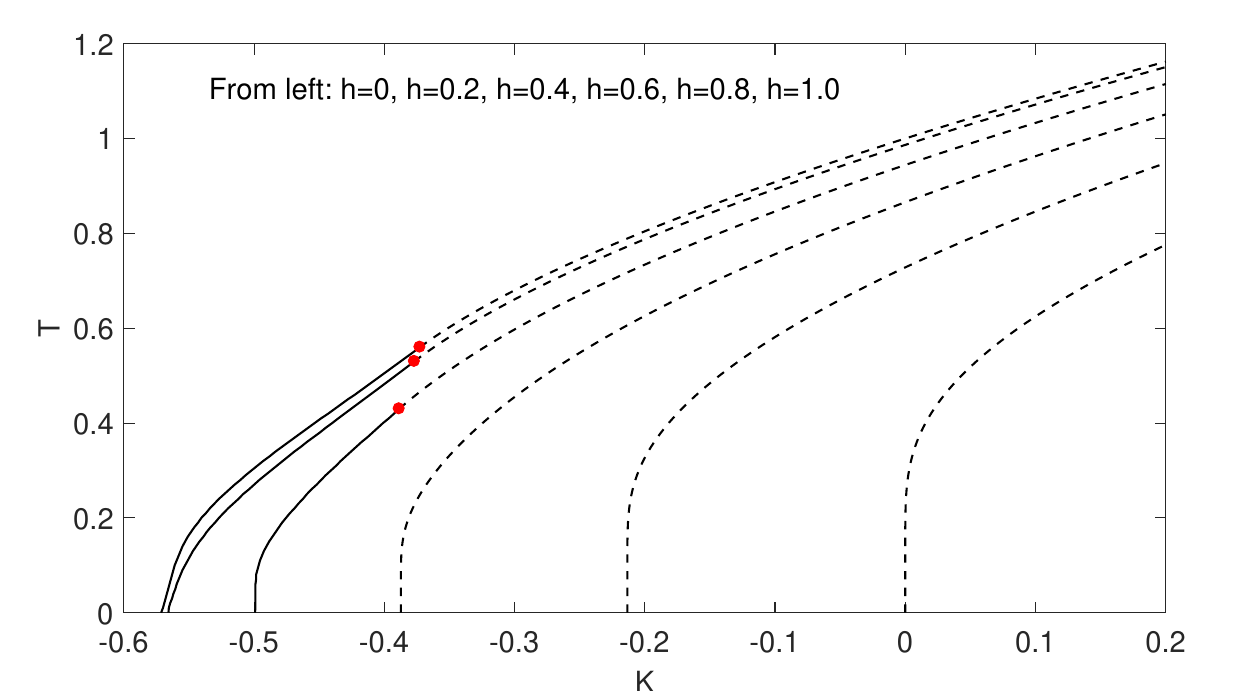}
\caption {The $(K,T)$ phase diagram for $p=8$. The different lines, corresponding to different values of the transverse field $h$ as indicated inside the plot, are the
phase transition lines as obtained from Eq. (\ref{free_en_p}). The full portions of the lines mark a first-order transition, while the dashed portions mark a
second-order transition. For the curves having both types of phase transitions, $h=0$, $h=0.2$ and $h=0.4$, the two portions meet at the tricritical point
denoted with a small red circle. The phase transition is represented by the passage from a vanishing (on the left of the curve) to a positive (on the right)
value of the magnetization $\langle \left( \sum S_i^z \right)/N \rangle$. The curves extend indefinitely to the right, as second-order lines, with an asymptotic value
for $T$, as explained below in the text, equal to $p$.}
\label{fig2}
\end{center}
\end{figure}

We have performed computations for $p=6$, $8$, $10$ and $12$. We present plots concerning the last three values of $p$. In the first part of this Section we show the results
obtained at finite $p$, while in the second part we will discuss the original model $p=\infty$.
In Fig. \ref{fig2} we show the most relevant
region of the $(K,T)$ phase diagram obtained for $p=8$.
The plot shows the phase transition lines, obtained from Eq. (\ref{free_en_p}) for various values of the transverse field $h$. The full portion of each curve corresponds
to points of first-order transition, while the dashed portion to points of second-order phase transition. On the left of the curve, for the corresponding value of $h$,
the system is disordered, i.e., the magnetization $m=\langle \left( \sum S_i^z \right)/N \rangle$ vanishes, while on the right the magnetization is positive; where the
transition is first-order the magnetization has a jump, when the transition is second-order it becomes positive continuously. We see that first-order transitions are
present, in the curves of the figure, for the smaller values of the transverse field $h$, while for the larger values the transition is always second-order.
For $h=0$ it is easily found analytically that at $T=0$ the phase transition line starts at $K=-p/[2(p-1)]$, value that tends to $-0.5$ for $p\to \infty$, coherently
with the results summarized above for the classical limit of the model. In the Introduction we have reminded the basic fact that the presence, in a system with
long-range interactions, of a first-order transition in the canonical ensemble, implies ensemble inequivalence with the microcanonical ensemble. We thus see, from
Fig. \ref{fig2}, that for small values of the transverse field $h$ the phase diagram maintains a region in which ensembles are not equivalent, but that at larger values
of $h$ there are only second-order transitions, so that we have ensemble equivalence. We will come back to this point after discussing the relevance of our results for the
original model, $p=\infty$.
Not visible in the range of $K$ shown in the plot,
we mention that the curves extend indefinitely to the right, as second-order lines, with an asymptotic value for $T$ equal to $p$. This can be understood
with the heuristic argument given in Appendix \ref{app:large}; however, for $h=0$ this can also be obtained analytically, see Eq. (\ref{Tcp}).

\begin{figure}[htbp]
\begin{center}
 \includegraphics[clip, trim=0cm 9cm 0cm 9cm, width=0.70\textwidth]{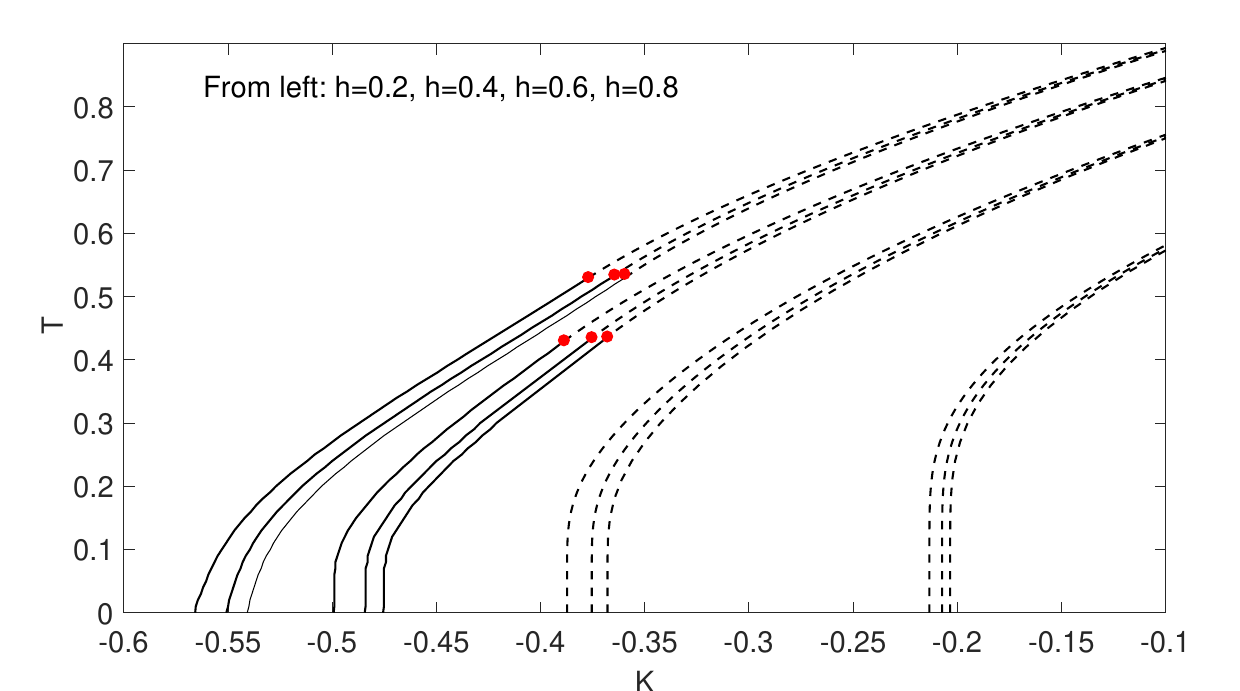}
\caption {The $(K,T)$ phase diagram for $p=8$, $p=10$ and $p=12$. Each one of the four bunches of three phase transition lines correspond to a value of the transverse field
$h$ as indicated in the plot. Inside each bunch, we have from left to right the phase transition line corresponding, respectively, to $p=8$, $p=10$ and $p=12$. On the left
of the curve, for the corresponding value of $p$ and $h$, the system is disordered, while it is magnetized on the right. For all three
values of $p$ the phase transition lines have a first-order portion (full line) and a second-order portion (dashed line) for the smaller values of $h$, i.e., $h=0.2$
and $h=0.4$, and only a second-order portion for the higher values of $h$. The small red circles are the tricritical points.}
\label{fig3}
\end{center}
\end{figure}

In the next plot, we show in Fig. \ref{fig3} the phase diagram for three different values of $p$, i.e., $p=8$, $p=10$ and $p=12$. For each value of $p$ there are the
phase transition lines for four different values of the transverse field $h$. The curves related to $p=8$ are obviously the same as those in Fig. \ref{fig2}.
In order to avoid overburdening the figure, with respect to Fig. \ref{fig2} we do not show the smallest and the largest value of $h$, i.e., $h=0$ and $h=1$ respectively,
and we show a smaller range of $K$.
Thus, to complete the information conveyed by this figure, we add that also for $p=10$ and $p=12$ the phase transition curves for $h=1$ start at the value $K=0$.
The curves appear in four bunches of three curves; each bunch corresponds to a value of $h$, while inside each bunch the curves correspond, from left to right, to $p=8$, $p=10$
and $p=12$. To have a convenient scale, the plot does not include the value $K=0$, where for a given value of $h$ the curves for different values of $p$ meet, as expected
(the cut of the nearest-neighbor interaction is irrelevant when the interaction is not present, as for $K=0$). As already mentioned, the curves
extend indefinitely to the right, with an asymptotic value in $T$ equal to the value of $p$ associated to the curve.

The information gathered up to now can be summarized as follows. {\it i}) For each $p$ the phase transition line starts at $T=0$ at a value of $K$ that increases with $h$;
more precisely, for $0\le h \le 1$, as shown in the plots, it is a negative value equal to $K=-p/[2(p-1)]$ for $h=0$ that decreases in modulus with increasing $h$.
For $h>1$, the starting point of the phase transition line keeps increasing, for each $p$, with $h$, starting however at a positive value of $K$. For $h=1$ the starting
point is $K=0$ for any $p$. This last property can be understood from the fact that for $K=0$ the results are independent from the value of $p$ and are equal to the results
pertaining to the LMG model, and as shown in Appendix \ref{app:lmg} the last model has a second-order quantum phase transition at $h=1$. {\it ii}) The phase transition lines have
a first-order portion for small values of $h$, while for larger values of $h$ the transition is second-order at any temperature. In the phase transition lines with a
first-order portion, this portion starts at $T=0$ and ends at a tricritical point. It is seen that the temperature of the tricritical point decreases at increasing $h$.
The first-order portions are contained in the region of negative $K$.

The last observation underlines the fact that the system shows a behavior similar to classical models: in order to have first-order transitions, and consequently
ensemble inequivalence, we need competition between the nearest-neighbor and the long-range interactions. Here we note that the introduction of a sufficiently large transverse
field removes the first-order portion, restoring ensemble equivalence. In agreement with this, we see that the first-order portion becomes smaller, with
the temperature of the tricritical point decreasing, by increasing $h$.

\begin{figure}[htbp]
\begin{center}
\includegraphics[clip, trim=0cm 8cm 0cm 9cm, width=0.70\textwidth]{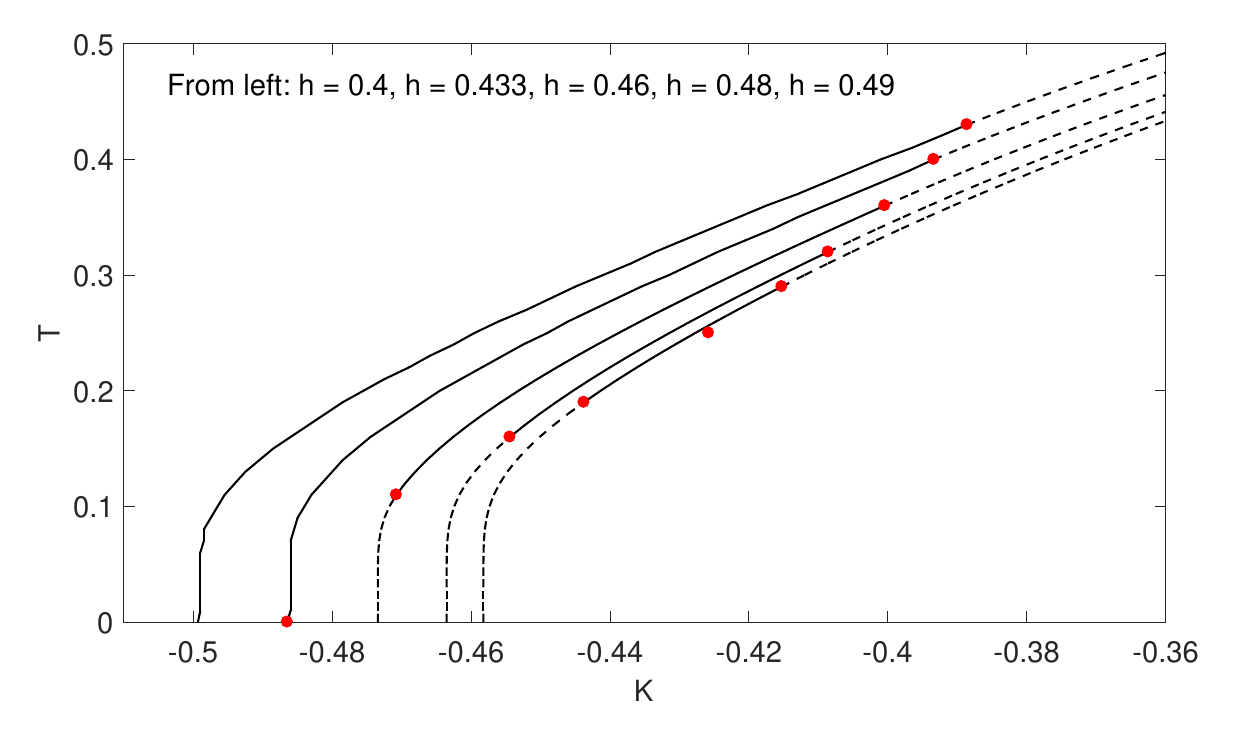}
\caption {The $(K,T)$ phase diagram for $p=8$ corresponding to $5$ different values of $h$ between $h=0.4$ and $h=0.49$. As in the previous plots, full (dashed) portions
correspond to first-order (second-order) transitions, and small red circles denote tricritical points. The phase transition lines corresponding to $h=0.46$, $h=0.48$ and
$h=0.49$ have a first-order portion between two finite temperatures, and included between one second-order portion starting at $T=0$ and ending at a lower tricritical point,
and another second-order portion starting at a upper tricritical point and extending indefinitely. The isolated red circles is the only one present in the phase transition
line (not plotted) corresponding to $h=0.495$.}
\label{fig4}
\end{center}
\end{figure}

\begin{figure}[htbp]
\begin{center}
 \includegraphics[clip, trim=0cm 9cm 0cm 9cm, width=0.70\textwidth]{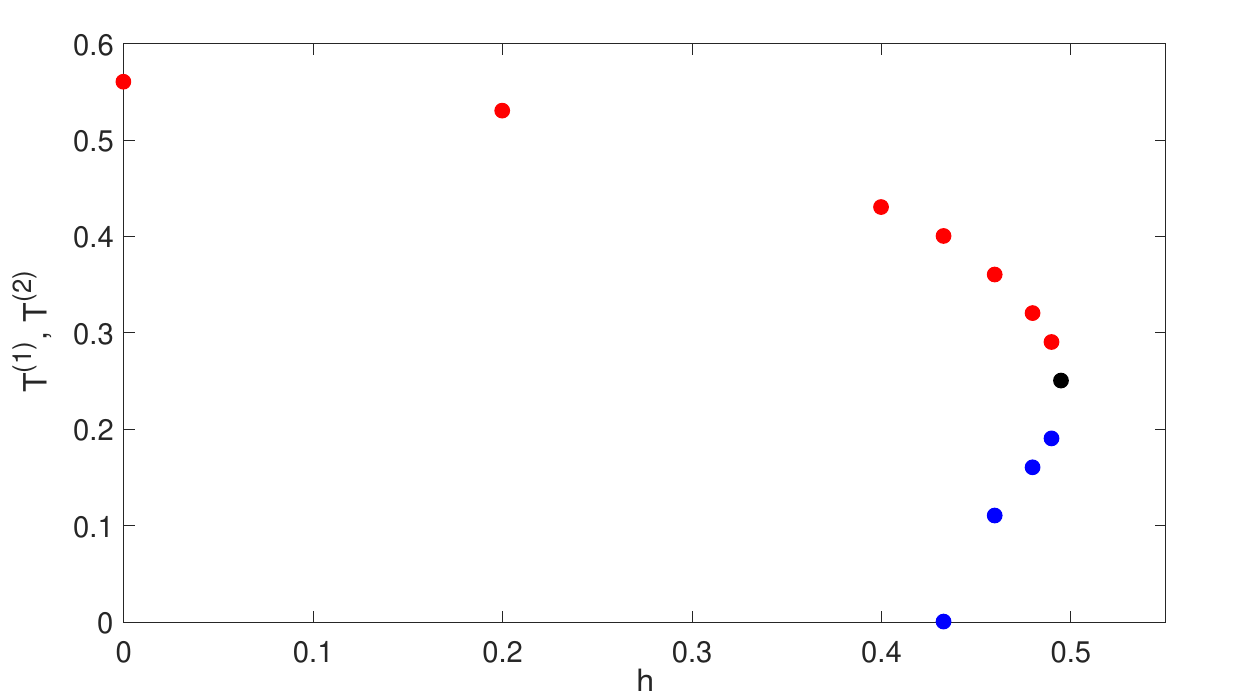}
\caption {The temperatures $T^{(1)}(h,p)$ (blue circles) and $T^{(2)}(h,p)$ (red circles) as a function of $h$ for $p=8$. The black circle denotes the common limit
of the two temperatures at $h=0.495$.}
\label{fig5}
\end{center}
\end{figure}

However, we have found an unexpected peculiarity. To explain it, we show in Fig. \ref{fig4} the phase transition lines, for $p=8$, for values of $h$ in a small range
between $h=0.4$ and $h=0.49$.
What we observe in the plot is the following. For three of the considered values of $h$, i.e., $h=0.46$, $h=0.48$ and $h=0.49$, the phase transition lines start with
a second-order portion at $T=0$; this portion ends at a lower tricritical point at finite temperature, after which there is a first-order portion
that ends at an upper tricritical point, after which the line is indefinitely second-order. The value $h=0.433$ is the one at which the lower tricritical point appears,
at $T=0$. For $h=0.495$ the two tricritical points coalesce, so that the
phase transition line is only second-order, as for all the larger values of $h$. A similar picture occurs also for the other values of $p$, although for a slightly different
range of $h$. Let us call $h_c(p)$ and $\widetilde{h}_c(p)$ the values of the transverse field, functions of $p$, at which the lower tricritical point appears at $T=0$
and at which the lower and the upper tricritical points coalesce, respectively. Also, we call $T^{(1)}(h,p)$ and $T^{(2)}(h,p)$ the values of the temperatures of the lower and
upper tricritical points, respectively, which are functions of $h$ and $p$. Finally, we call $\widetilde{T}(p)$ the common limit of
$T^{(1)}(h,p)$ and $T^{(2)}(h,p)$ when $h \to \widetilde{h}_c(p)$. In Table \ref{tab1} we give the values of $h_c(p)$, $\widetilde{h}_c(p)$ and of $\widetilde{T}(p)$
for the different values of $p$.
\begin{table}[ht]
\begin{center}
\begin{tabular}{|p{1cm}|p{1.5cm}|p{1.5cm}|p{1.5cm}|}
\hline
\makebox[1cm][c]{$p$} & \makebox[1.5cm][c]{$h_c$} & \makebox[1.5cm][c]{$\widetilde{h}_c$} & \makebox[1.5cm][c]{$\widetilde{T}$} \\
\hline
\makebox[1cm][c]{$6$} & \makebox[1.5cm][c]{$0.425$} & \makebox[1.5cm][c]{$0.485$} & \makebox[1.5cm][c]{$0.260$} \\
\hline
\makebox[1cm][c]{$8$} & \makebox[1.5cm][c]{$0.433$} & \makebox[1.5cm][c]{$0.495$} & \makebox[1.5cm][c]{$0.250$} \\
\hline
\makebox[1cm][c]{$10$} & \makebox[1.5cm][c]{$0.459$} & \makebox[1.5cm][c]{$0.505$} & \makebox[1.5cm][c]{$0.235$} \\
\hline
\makebox[1cm][c]{$12$} & \makebox[1.5cm][c]{$0.475$} & \makebox[1.5cm][c]{$0.509$} & \makebox[1.5cm][c]{$0.230$} \\
\hline
\end{tabular}
\caption{The values of $h_c$, $\widetilde{h}_c$ and $\widetilde{T}$ at which the lower tricritical point appears, the lower and the upper tricritical
points coalesce, and the temperature of the coalescence, respectively.}
\label{tab1}
\end{center}
\end{table}
To see the behavior of the temperatures $T^{(1)}(h,p)$ and $T^{(2)}(h,p)$ we show in Fig. \ref{fig5} their plot as a function of $h$ for $p=8$.
In the following, discussing the original model, $p=\infty$, we will comment on the relevance of this peculiarity.

\subsection{Rescaling {\bf of the coupling $K$} and the phase diagram for the quantum Nagle-Kardar model}
\label{subsec:rescal}
The value $p=12$ is the largest that we could analyze in a reasonable computing time (that essentially depends on the diagonalization of $2^p \times 2^p$ matrices).
Although such a value could seem still too small to have a reliable estimate of the phase diagram of the quantum Nagke-Kardar model, corresponding to $p=\infty$,
we will now argue that it is nevertheless possible to establish such estimate.

The classical model, $h=0$, as already pointed out, can be studied analytically for any value of $p$. Not only it is easy to compute that the phase transition
line starts at $T=0$ at the value $K=-p/[2(p-1)]$, as already remarked (retrieving the value $K=-1/2$ of the classical Nagle-Kardar model
for $p \to \infty$ \cite{kardar1983crossover}), but one can also obtain an analytical expression for the second-order phase transition line, giving the critical
temperature $T_c$, or equivalently its inverse $\beta_c$, as a function of $K$. It is given implicitly, for $K$ larger than its negative value at the tricritical point,
by
\bea
&&p \left( \beta_c \er^{\beta_c K}-1\right) +
\frac{\beta_c \left( 1-\er^{2\beta_c K}\right)}{2} \nonumber \\
&+&\frac{\beta_c \left( 1-\er^{\beta_c K}\right)^2}{2} 
\tanh^{p-1}{\left( \frac{\beta_c K}{2}\right)} = 0\, .
\label{Tcp}
\eea
For $p\to\infty$ one gets $K=T_c \log{T_c}$, while for $K\to \infty$ one gets $T_c=p$. We proceed as follows. From the last expression we derive an expression for
$T_c$ at small absolute values of $K$. At first order in $K$ we get
\begin{equation}
\label{tcrescal}
T_c=1+\frac{p-1}{p} K \, .
\end{equation}
This implies that, if for a small value of $K$ we have a certain value $T_c$ for $p\to \infty$, then for a finite $p$ we would get the same
critical temperature $T_c$ with the rescaling 
\begin{equation}
K \to \widetilde {K}\equiv \frac{p}{p-1} K \, .
\label{collapse}
\end{equation}
This can be understood physically with the following reasoning. Performing a $p$-blocking means removing one every $p$ nearest-neighbor interactions; therefore one could
say that averaging over the nearest-neighbor couples the coupling coefficient is $(p-1)/p$ times the one appearing in the Hamiltonian, and thus the results
one would get are those pertaining to the original system with the coupling coefficient rescaled in this way. This reasoning can be applied for small absolute values
of $K$, since it is not guaranteed that the scaling would work at larger values.

\begin{figure}[htbp]
\begin{center}
\includegraphics[clip, trim=0cm 9cm 0cm 9cm, width=0.70\textwidth]{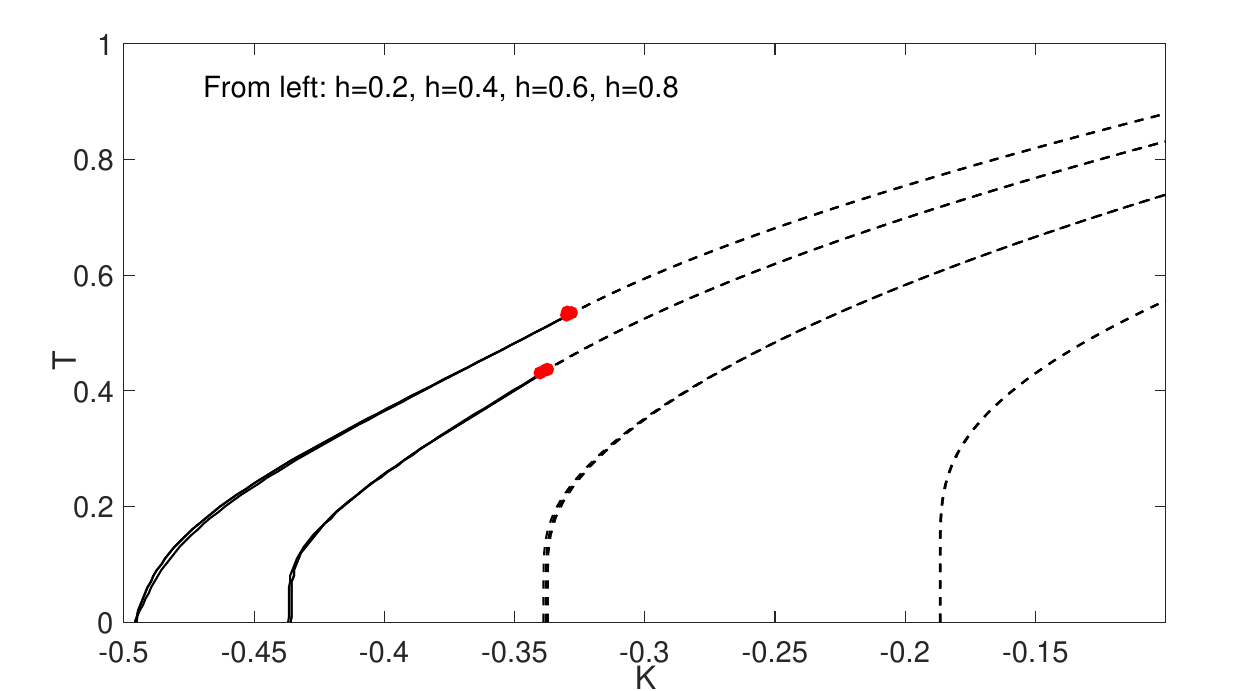}
\caption {The $(K,T)$ phase diagram for $p=8$, $p=10$ and $p=12$. The curves are the same as those in fig. \ref{fig3}, but in which the scaling $K \to [(p-1)/p]K$
has been performed. For each value of $h$ the curves for the different values of $p$ superimpose almost perfectly. The blurry appearance of each of the four bunches
of three different curves has been intentionally left. Also the tricritical points for $h=0.2$ and $h=0.4$, denoted by the red circles, superimpose almost perfectly.
As in the previous figures, full (dashed) portions of the curves correspond to first-order (second-order) transitions.}
\label{fig6}
\end{center}
\end{figure}

Nevertheless let us see what happens by performing this scaling to the region of the phase diagram shown in Fig. \ref{fig2} and Fig. \ref{fig3}.
We show in Fig. \ref{fig6} the same curves plotted in Fig. \ref{fig3}, but in which we have performed the scaling $K \to [(p-1)/p]K$.
We see that for each value of $h$ the curves corresponding to the different values of $p$ collapse almost perfectly on the same curve. From this fact we can infer
the following. Although the above scaling has been obtained with a computation only at first order in $K$ and for $h=0$, and furthermore only for the second-order
portion of the phase transition line, it seems that it works very good also for values of $K$ that, despite being less than $1$ in absolute value, are not
very small, and above all also when the transverse field $h$ is different from zero. It is important to note that also the tricritical points superimpose almost perfectly,
and this without having made any rescaling in the temperature.
We emphasize that the scaling cannot work for any value of $K$, as can be
understood from the simple fact that, when $K$ becomes large and positive, the phase transition line for different values of $p$ tends to a temperature equal to $p$,
and therefore the different curves cannot superimpose. As a matter of fact we have checked that when $K$ is positive and larger than $1$ the rescaled curves
begin to separate. Fortunately, the most interesting region of the phase diagram is in the range of $K$ shown in the plot
of Fig. \ref{fig6}; it is the region where the first-order portions of the phase transition lines occur, and therefore the region where ensemble inequivalence
is located.

From this collapse of the curves we can reasonably infer that the rescaled curves represent the phase transition lines for the original quantum Nagle-Kardar model,
{\it as far as the region at small absolute values of $K$ is concerned}; region, we emphasize again, where ensemble inequivalence occurs. In other words, the curves in
Fig. \ref{fig6} can be taken to represent, for that range of $K$, the phase transition lines of the quantum Nagle-Kardar model, including the position of the tricritical
points for the various values of $h$. However, to give more solidity to this conclusion requires that we face the problem of the peculiar property analyzed above,
related to the existence of the small range of $h$, for each $p$, where the phase transition lines present two second order portions, as shown in Fig. \ref{fig4}.
In practice, we should determine  whether the existence of two different critical values of $h$, i.e., $h_c(p)$ and $\widetilde{h}_c(p)$ is an artifact of
the $p$-blocking, i.e., of having a finite value of $p$. We note, in particular, that considering a value of $h$ which is between $h_c$ and $\widetilde{h}_c$ for a
given value of $p$, but outside this range for another value of $p$, something that it is possible, by looking at the values on Table \ref{tab1}, it is clear that, even
if the two phase transition lines superimpose after the $K$ rescaling, one of the two lines would still have two tricritical points, contrary to the other.
Then, with the purpose to shed light on the problem raised by this peculiarity, we have used the available values of these critical values of $h$, as given in
Table \ref{tab1}, to perform a fit, trying to see what one should expect at large values of $p$ for $h_c$ and $\widetilde{h}_c$. Actually, we have performed two
different fits, using two different trial functions. The first trial function is $a+b\er^{-cp}$. For $h_c$ we have obtained the following values:
$a=0.49 \pm 0.01$, $b=-0.10 \pm 0.03$, $c=0.06 \pm 0.02$. For $\widetilde{h}_c$ we found: $a=0.513 \pm 0.002$, $b=-0.06 \pm 0.02$, $c=0.35 \pm 0.10$. As second trial
function we used $a+b/p^c$. For $h_c$ we obtained: $a=0.50 \pm 0.02$, $b=-0.35 \pm 0.10$, $c=1.2 \pm 0.4$. For $\widetilde{h}_c$ we found: $a=0.51 \pm 0.01$,
$b=-0.20 \pm 0.05$, $c=1.0 \pm 0.3$. Of course the most important parameter is $a$, that denotes the asymptotic value for $p\to \infty$. From the comparison of the two
fits we can conservatively say that the limits of the two critical values of $h$ can be taken to be: $h_c(p\to \infty) \approx 0.50 \pm 0.02$ and
$\widetilde{h}_c(p\to \infty) \approx 0.51 \pm 0.01$. Although considering the errors one could say that for large $p$ the two values $h_c$ and $\widetilde{h}_c$
coincide, this cannot be taken as a proven fact.

In conclusion of our analysis, we think that one can say that we have determined the phase transition lines of the quantum Nagle-Kardar model in the most
relevant region of the $(K,T)$ phase diagram, the region where first-order transitions appear and consequently ensemble inequivalence occurs. In doing this
we have also determined when the transition is first-order and when it is second-order, finding at the same time the location of the tricritical points.
However it is fair to say also that there is a small range of $h$, around $0.5$, where it is not clear if the peculiarity found in our computations at finite $p$,
i.e., the presence of a small range in $h$ in which the phase transition lines have two second-order portions with a first-order portion between them, and two tricritical
points, survives for $p \to \infty$ for an even smaller range of $h$. However, the main purpose of our analysis is independent from this fact: we have 
shown that in the quantum Nagle-Kardar model there is a critical value of $h$ when ensemble equivalence, absent in the classical model, is restored. In practice,
beyond the critical value of $h$ quantum fluctuations play the role that in the classical case is played by a sufficiently high temperature, and therefore the
mean-field ferromagnetic interaction is not sufficient, even at small temperatures, to cause a first-order transition.



\section{Discussion and Conclusions}
\label{sec:concl}
In this work we have studied the quantum Nagle-Kardar model, obtained by adding to the classical Nagle-Kardar model a transverse magnetic field $h$.
The main purpose was to determine the role of the quantum fluctuations in relations to the presence or not of ensemble inequivalence between
the canonical and the microcanonical ensemble. In this respect, we have relied on the established fact that the presence of a first-order phase transition
in the canonical ensemble implies, in a system with long-range interactions, ensemble inequivalence, due to the absence of phase separation.

From the technical point of view we had to face the problem that, by introducing a transverse field, and therefore having a Hamiltonian in which there
are non-commuting operators, the Hubbard-Stratonovich transformation, which is the main tool used in the solution of the classical model, is not
exact in the quantum case. Nevertheless, we could provide an argument to show that, {\it in the thermodynamic limit} the problem caused by the non-commutativity
can be overcome. This could not yet allow an easy computation of the partition function, since the presence of the nearest-neighbor interaction,
that in the classical case is tackled with a transfer matrix method, cannot be dealt with using the same method, since the transfer matrix would be of the same
size of the Hamiltonian operator, and therefore useless in the thermodynamic limit. Then we resorted to what we have called $p$-blocking; we have studied a system
in which the nearest-neighbor interaction is removed in one every $p$ couples. This allowed to compute the partition function in the thermodynamic limit, at the
price of working with $2^p \times 2^p$ matrices. Although this size permits to make computations for values of $p$ not very large (we could go up to $p=12$),
we could see that a simple rescaling in the plots of the phase diagram for finite values of $p$ allows to obtain the phase diagram of the full model ($p=\infty$).
More precisely, we could determine the phase diagram in the most relevant region, that at small absolute values of the nearest-neighbor coupling $K$. This
was sufficient to determine when ensemble inequivalence occurs and when it is restored by the quantum fluctuations. In this analysis we found a peculiar behavior
which is present for a small range of values of $h$: in this range there is a (possibly spurious) first-order line at finite temperature between two second-order lines, one
of them originating at $T=0$. We argued that this range should shrink to zero in the limit $p \to \infty$, but we could not determine this with certainty. We
have anyway  clarified that this point does not interfere with the main purpose of the study, i.e., the determination of a value of the transverse field, around $h=0.5$
(or $h=0.5J$ if we put back the mean-field coupling coefficient different from $1$), where ensemble equivalence is definitely restored at all temperatures.
This means that a sufficiently high transverse field, or in other words a sufficiently high level of quantum fluctuations prevents the possibility that the competition
between long-range and short-range interactions induces ensemble inequivalence.

We have studied the two-dimensional $(K,T)$ phase diagram for different values of the transverse field $h$. Obviously, from the technical point of view one
could consider $h$ on the same footing of the other Hamiltonian parameter, $K$, and build a three-dimensional $(K,h,T)$ phase diagram. However, this
would not change our conclusions concerning the role of the transverse field $h$ on the equivalence or inequivalence of ensembles.

We point out that the argument showing that in the thermodynamic limit the non-commutativity of the operators appearing in the Hamiltonian can be neglected,
as far as the Hubbard-Stratonovich transformation is concerned, is not limited to this specific model. In fact, in the argument we used only the fact that the operators
$S_i^z$ do not commute with all the operators appearing in the local part of the Hamiltonian (i.e., the part not related to the mean-field interaction).
Therefore, the argument could be used for all Hamiltonians regardless of the operators appearing in the local part. As a byproduct, this implies that the partition function
of a system of this type, always in the thermodynamic limit, is equal to the partition function of the system in which the mean-field interaction is replaced
by an interaction with an external degree of freedom, with a coupling coefficient scaling as $1/\sqrt{N}$. For models without a nearest-neighbor interaction,
as in the LMG model, this was already argued in, e.g., Ref. \cite{romanroche2023exact}; our argument can extend this correspondence to a larger class of models, since any
local interaction could be included.

We stress that the peculiarity found in our computations, i.e., the presence, in the systems with finite $p$, of a first-order portion in the phase transition line
between two second-order portions, although not so common, cannot be considered a 
specific thing appearing in just this model. For example, something similar
is found also in classical models with both mean-field and short-range interactions \cite{campa2019ising}, for certain values of the Hamiltonian parameters. This fact
has also an implication that should be taken into account in this kind of studies. Although there is a general tendency for first-order transitions to appear at lower
temperatures than second-order transitions, this is not automatic; sufficiently complex Hamiltonians can give rise to thermodynamic phase diagrams in which there are
second-order transitions at lower temperatures than first-order transitions.

Although in this work we have been concerned only with equilibrium properties, we find it useful to remark that the occurrence of ensemble inequivalence has direct interest,
for both classical and quantum systems, also for the problem of approach to equilibrium, since the equilibrium state reached by the system will be the one related
to the control variables employed in its preparation, e.g., fixed energy of fixed temperature.

In conclusion, we have shown that the addition of quantum fluctuations to a classical model restore ensemble equivalence; in our case the classical Nagle-Kardar model
has been transformed to a quantum model by adding a transverse field. Probably also this fact cannot be taken as automatically verified in all models. However,
it is likely to be found in many cases; in fact, on physical grounds one could consider quantum fluctuations as playing the role of additional thermal fluctuations
since they create new eigenstates of the quantum  Hamiltonian by mixing, in general, those of the classical Hamiltonian. It would be valuable to find a
model-independent criterion to see when this happens.


\section*{Acknowledgments}

Discussions with Giovanna Morigi and correspondence with Alessio Lerose are gratefully acknowledged. A.C. acknowledges hospitality from the Department
of Physics of the University of Trieste during the last stages of the preparation of the manuscript.

\appendix

\section{Bethe-Peierls approximation for the classical Nagle-Kardar}
\label{app:bethe}

We describe here the Bethe-Peierls approximation for the classical case, i.e., for the model represented by the Hamiltonian (\ref{origham}) in which the transverse
field vanishes, $h=0$. To begin with, however,
we consider the classical Ising model in an external magnetic field $B$, namely the Hamiltonian
\be
H = -\frac{K}{2} \sum_i S_i S_{i+1} - B \sum_i S_i \, .
\ee
with the understanding that we are working on the $z$ component of the spin, and we are adopting periodic boundary conditions. 
This model is analytically solvable, and in particular the average magnetization $\langle S \rangle$, site-independent in the thermodynamic limit,
at the inverse temperature $\beta$ is given by
\be
\label{exacts}
\langle S \rangle = \frac{\sinh \beta B}{\left[\sinh^2 \beta B + \er^{-2\beta K}\right]^{\frac{1}{2}}}
\ee
The Bethe-Peierls approximation consists in studying the three-sites
system (the spins being indexed by $\left\{ S_{-1},S_0,S_{+1}\right\}$) with Hamiltonian
\be
H = -B S_0 -\frac{K}{2} \left( S_{-1}S_0 + S_0 S_{+1}\right) -B' \left( S_{-1}+S_{+1}\right)
\ee
The difference of the external magnetic field on the two spins $S_{-1}$ and $S_{+1}$, i.e., the quantity $B'-B$, is meant to summarize the interactions
(direct and mediated) of these two spins with all the other spins that are not explicitly considered in this reduced system. One then computes the averages
$\langle S_0\rangle$ and $\frac{1}{2}\langle S_{-1} + S_{+1}\rangle$; imposing the condition that they are equal fixes the value of $B'$.
At the end one finds that the average magnetization $\langle S_0\rangle$ is equal to the exact expression (\ref{exacts}).
In more details, one finds:
\bea
\langle S_0 \rangle &=& \frac{\er^{\beta B}\cosh^2 a - \er^{-\beta B}\cosh^2 b}
{\er^{\beta B}\cosh^2 a + \er^{-\beta B}\cosh^2 b} \nonumber \\
\frac{\langle S_{+1}+S_{-1}\rangle}{2} &=&
\frac{\er^{\beta B}\cosh^2 a\tanh a + \er^{-\beta B}\cosh^2 b\tanh b}
{\er^{\beta B}\cosh^2 a + \er^{-\beta B}\cosh^2 b} \nonumber \\
\er^{2\beta B' -2\beta B} &=& \frac{\cosh a}{\cosh b} \, ,
\label{eqbp}
\eea
where $a=\beta B +\beta K/2$ and $b=\beta B -\beta K/2$. Using the third expression in the first we have
\be
\label{s0bb}
\langle S_0 \rangle = \frac{\er^{4\beta B'}-\er^{2\beta B}}{\er^{4\beta B'}+\er^{2\beta B}}
\ee
The third expression in (\ref{eqbp}) can also be used to have an expression of $\er^{\beta B'}$ as a function of $\beta B$ and $\beta K$. In fact, one gets
the equation
\be
\er^{4\beta B'}-\er^{\beta K}(\er^{2\beta B}-1)\er^{2\beta B'}-\er^{2\beta B}=0
\ee
from which
\be
\er^{2\beta B'} = \frac{\er^{\beta K}(\er^{2\beta B}-1)+ \sqrt{\er^{2\beta K}(\er^{2\beta B}-1)^2+4\er^{2\beta B}}}{2}
\ee
Substituting in the equation (\ref{s0bb}) for $\langle S_0 \rangle$ we find, after straightforward passages
\be
\langle S_0 \rangle = \frac{\sinh(\beta B)}{\sqrt{\sinh^2(\beta B)+\er^{-2\beta K}}}
\ee
which is the exact expression, Eq. (\ref{exacts}). In this procedure there was one variable to be determined, i.e. $B'$, that was obtained by equalizing
$\langle S_0 \rangle$ and $(\langle S_{+1}+S_{-1}\rangle)/2$.

Going now to the classical Ising model with the mean-field interaction, i.e., the Nagle-Kardar model, with Hamiltonian
\be
H = -\frac{J}{2N}\sum_{i,j} S_i S_j -\frac{K}{2} \sum_i S_i S_{i+1} \, ,
\ee
the partition function and then the average magnetization $m \equiv \langle S \rangle$ at inverse temperature $\beta$ can be computed with the help
of the Hubbard-Stratonovich transformation, and one obtains that $m$ is given by the solution of the self-consistent equation \cite{campa2019ising}
\be\label{exactmeanb}
m = \frac{\sinh \beta J m}{\left[\sinh^2 \beta J m + \er^{-2\beta K}\right]^{\frac{1}{2}}}
\ee
In the Bethe-Peierls approach this can be obtained from the above procedure for the chain in an external field $B$ in the following way.
Now the three-sites Hamiltonian is
\be
H = -Jm S_0 -\frac{K}{2} \left( S_{-1}S_0 + S_0 S_{+1}\right) -B' \left( S_{-1}+S_{+1}\right)
\ee
where the mean-field $Jm$ substitutes the external field $B$. Then, now there are two variables to be determined, namely $B'$ and $m$. Therefore, now we
need two equations. The first, like before, is $\langle S_0 \rangle = (\langle S_{+1}+S_{-1}\rangle)/2$;  the second is simply that both members are equal to $m$. From the above
calculations we have, substituting $B$ with $Jm$,
\be
\langle S_0 \rangle = \frac{\sinh(\beta Jm)}{\sqrt{\sinh^2(\beta Jm)+\er^{-2\beta K}}}
\ee
Putting this equal to $m$ we obtain the exact self-consistent equation (\ref{exactmeanb}). Therefore for the classical Nagle-Kardar model the Bethe-Peierls
approximation gives the exact result.

\section{The Gaussian integral involving a scalar and an operator}
\label{app:gauss}

Suppose to have the integral
\be
\label{appc1}
\int \dd x \, \exp \left[-\frac{N}{2}x^2 + xO \right] \, ,
\ee
with $O$ a symmetric operator. We know that this integral is equal to the operator
\be
\label{appc2}
\sqrt{\frac{2\pi}{N}} \exp \left[ \frac{O^2}{2N}\right] \, .
\ee
The meaning of the loose statement that the last exponential operator is equal to that obtained by
putting $x=O/N$ in the exponent in the integrand in (\ref{appc1}) (and that when $N$ is very large
this is also the dominant term in the integral) can be formalized with the procedure by which
it is seen that Eq. (\ref{appc2}) is equal to Eq. (\ref{appc1}). In fact, if $U$ is the unitary operator
that diagonalizes $O$, so that $U^\dag O U = D$, with $D$ diagonal, Eq. (\ref{appc1}) can be written as
\be
\label{appc3}
U \left( \int \dd x \, \exp \left[ -\frac{N}{2}x^2 +xD \right] \right) U^\dag
\ee
The operator in the integrand is diagonal, and for the $i$-th diagonal element the integral is equal
to
\be
\label{appc4}
\sqrt{(2\pi)/N}\exp[D_{ii}^2/(2N)] \, .
\ee
The expression in the exponent is the one we would get by
putting $x=D_{ii}/N$ in the expression $[-(Nx^2)/2 + x D_{ii}]$. It is the equality $x=D_{ii}/N$, for each
diagonal element of operator $O$ after its diagonalization, that one refers to with the loose expression
$x=O/N$.

The passages in this appendix are only to give a more formal meaning to the last expression, but they are not
relevant for the actual computation of the partition function and the free energy, where
the saddle point evaluation is performed for the integral of a scalar function
(see Eq. (\ref{free_en_p})), since the purpose to introduce the HS transformation is to compute
the integration over the auxiliary variable $x$ after the computation of the trace over the spin variables.

\section{The free energy for the LMG model}
\label{app:lmg}
When $K=0$ the effective Hamiltonian reduces to
\be
\label{tildehamK0}
\widetilde{H}(x) = -h \sum_{i} S_{i}^x -\sqrt{J}x \sum_{i} S_{i}^z \, .
\ee
In this case in the partition function in Eq. (\ref{parths}) the spins decouple, and performing the trace one obtains
\be
\label{parthsK0}
Z=\sqrt{\frac{\beta N}{2 \pi}} 
\int \dd x \, \er^{-N \left[\frac{\beta}{2}x^2 -\ln \left( 2\cosh \left[\beta \sqrt{x^2+h^2}\right] \right)
\right]} \, .
\ee
Performing the integral, in the thermodynamic limit, with the saddle point one finds the following results for the
magnetization $m=\langle \left( \sum S_i^z \right)/N \rangle$. For $h>1$ the system is always disordered, i.e., it
is $m=0$ for any temperature. On the other hand, for $h<1$ there is a critical value of the inverse temperature, $\beta_c$
where the system has a second-order phase transition from an ordered to a disordered state. For $\beta < \beta_c$ the system
is paramagnetic ($m=0$), while for $\beta>\beta_c$ the magnetization $m$ is given by the solution of the equation
\be
\label{mlmg}
\sqrt{m^2+h^2} = \tanh \left( \beta \sqrt{m^2+h^2} \right) \, .
\ee
The critical temperature $T_c=1/\beta_c$ is given by
\be
\label{tcritlmg}
T_c = 2h \left[ \ln \left( \frac{1+h}{1-h}\right)\right]^{-1} \, .
\ee
$T_c$ decreases from $T_c=1$ for $h=0$ to $T_c=0$ for $h=1$ (the latter denoting the second-order quantum
phase transition at vanishing temperature). For a given $h\le 1$ the magnetization $m$ decreases from
$\sqrt{1-h^2}$ at $T=0$ to $0$ at $T=T_c$.

\section{The phase transition lines for large nearest-neighbor interaction}
\label{app:large}

For positive very large $K$ it is possible to show that the system with a $p$-blocking has a second-order transition at $T=p$.
Let us begin with the classical case, assuming that in the $p$-blocking Hamiltonian defined by Eqs. (\ref{pblockham})
and (\ref{Hcut}) $h$ is put equal to zero and $K$ is positive very large. One can argue that in each block the $p$ spins
will be aligned at each finite temperature, and that the system will be equivalent to a system with only the mean-field
interaction with Hamiltonian
\be
H = -\frac{Jp}{2M} \sum_{i,j} S_i^z S_j^z \, ,
\label{orighamklarge}
\ee
where $M=N/p$, and where in the sum $i$ and $j$ go from $1$ to $M$. In fact, the aligned spins in each block will form a single spin that
can have values $+p$ and $-p$. Dividing each such spin by $p$, in order to normalize it to values $+1$ and $-1$, we have to multiply
the coupling constant by $p^2$. One power of $p$ is used to go from $N$ to $N/p$, so that we obtain the Hamiltonian (\ref{orighamklarge}), where now
$M$ plays the role of the number of spin. We know, as mentioned in the main text when discussing the various limits of our model, that
in this case we have a second-order transition at a temperature equal to the coupling constant of the mean-field interaction, therefore
equal to $Jp$, or $p$ when we pose $J=1$.

When we introduce the transverse field $h$ the transition temperature for any $K$ will decrease by increasing $h$. However,
when $K\to \infty$ we can argue that, for any finite $h$, the system will behave as for $h=0$, thus having a second-order transition at $T=p$.

We note that this result is coherent with the fact that for the original system, $p\to \infty$, the transition temperature increases indefinitely
with $K$.

\bibliographystyle{apsrev4-2}
\bibliography{biblio}


\end{document}